\documentclass[twocolumn]{aastex631}

\usepackage{color}
\usepackage{url}
\usepackage{hyperref}
\usepackage{xspace}
\usepackage{soul}
\usepackage{hyperref}

\newcommand{\source}{{Swift~J1727.8$–$1613}\xspace}

\begin{document}

\title{Discovery of X-ray Polarization from the Black Hole Transient Swift~J1727.8$–$1613}

\author[0000-0002-5767-7253]{Alexandra Veledina}
\affiliation{Department of Physics and Astronomy, 20014 University of Turku, Finland}
\affiliation{Nordita, KTH Royal Institute of Technology and Stockholm University, Hannes Alfv\'ens v\"ag 12, SE-10691 Stockholm, Sweden}
\author[0000-0003-3331-3794]{Fabio Muleri}
\affiliation{INAF Istituto di Astrofisica e Planetologia Spaziali, Via del Fosso del Cavaliere 100, 00133 Roma, Italy}
\author[0000-0003-0079-1239]{Michal Dov\v{c}iak}
\affiliation{Astronomical Institute of the Czech Academy of Sciences, Bo\v{c}n\'{i} II 1401/1, 14100 Praha 4, Czech Republic}
\author[0000-0002-0983-0049]{Juri Poutanen}
\affiliation{Department of Physics and Astronomy,  20014 University of Turku, Finland}
\author[0000-0003-0411-4243]{Ajay Ratheesh}
\affiliation{INAF Istituto di Astrofisica e Planetologia Spaziali, Via del Fosso del Cavaliere 100, 00133 Roma, Italy}
\author[0000-0002-6384-3027]{Fiamma Capitanio}
\affiliation{INAF Istituto di Astrofisica e Planetologia Spaziali, Via del Fosso del Cavaliere 100, 00133 Roma, Italy}
\author[0000-0002-2152-0916]{Giorgio Matt}
\affiliation{Dipartimento di Matematica e Fisica, Universit\`{a} degli Studi Roma Tre, Via della Vasca Navale 84, 00146 Roma, Italy}
\author[0000-0002-7781-4104]{Paolo Soffitta}
\affiliation{INAF Istituto di Astrofisica e Planetologia Spaziali, Via del Fosso del Cavaliere 100, 00133 Roma, Italy}
\author[0000-0002-9443-6774]{Allyn F. Tennant}
\affiliation{NASA Marshall Space Flight Center, Huntsville, AL 35812, USA}
\author[0000-0002-6548-5622]{Michela Negro} 
\affiliation{Department of Physics and Astronomy, Louisiana State University, Baton Rouge, LA 70803, USA}
% ...


\begin{abstract}
We report the first detection of the X-ray polarization of the bright transient \source with the Imaging X-ray Polarimetry Explorer. 
The observation was performed at the beginning of the 2023 discovery outburst, when the source resided in the bright hard state.
We find a time- and energy-averaged polarization degree of $4.1\%\pm0.2\%$ and a polarization angle of $2\fdg2\pm1\fdg3$ (errors at 68\% confidence level; this translates to $\sim$20$\sigma$ significance of the polarization detection).
This finding suggests that the hot corona emitting the bulk of the detected X-rays is elongated, rather than spherical.
The X-ray polarization angle is consistent with that found in sub-mm wavelengths.
Since the sub-mm polarization was found to be aligned with the jet direction in other X-ray binaries, this indicates that the corona is elongated orthogonal to the jet.
\end{abstract}



\section{Introduction}
\label{sec:intro}

Mass accretion is a fundamental process of energy extraction that powers some of the brightest X-ray sources.
It operates very efficiently for compact objects, neutron stars and black holes (BHs), which accrete matter from a nearby companion star.
BH X-ray binaries display two major spectral states, hard and soft, which are thought to be linked to different accretion regimes \citep{Zdziarski2004,Remillard2006,Done2007}.
In the soft state, the X-ray spectrum is dominated by a blackbody-like emission that is believed to be produced by a geometrically thin, optically thick accretion disk \citep{SS73,NT73,Page1974}. 
In the hard state, the spectrum constitutes a power-law-like continuum, that is generally attributed to Comptonization in a medium with hot electrons (a corona).
At the transitions between the canonical states, the hard-intermediate and soft-intermediate, as well as very high (or steep power-law) states have been identified \citep{Homan2005,Belloni2010}.

Transitions of BH X-ray binaries between various spectral states follow a well-known pattern, however, our understanding of the accretion geometry and physical mechanisms producing broadband emission in these states is still incomplete.
The geometry and location of the hot medium in the hard spectral state is still a matter of debate \citep{Poutanen2018,Bambi2021}. 
The structure of the accretion disk is unclear, its stability in the soft state are puzzling \citep{Dexter2012,Jiang2013}. 
Furthermore, the conditions for the source to show the very high state are unknown \citep{Done2007}.
X-ray polarimetry is a new diagnostic that may help resolve questions regarding the geometry of the emission region left unanswered by the conventional tools of spectroscopy and timing.

\begin{figure*} 
\centering
\includegraphics[width=0.80\linewidth]{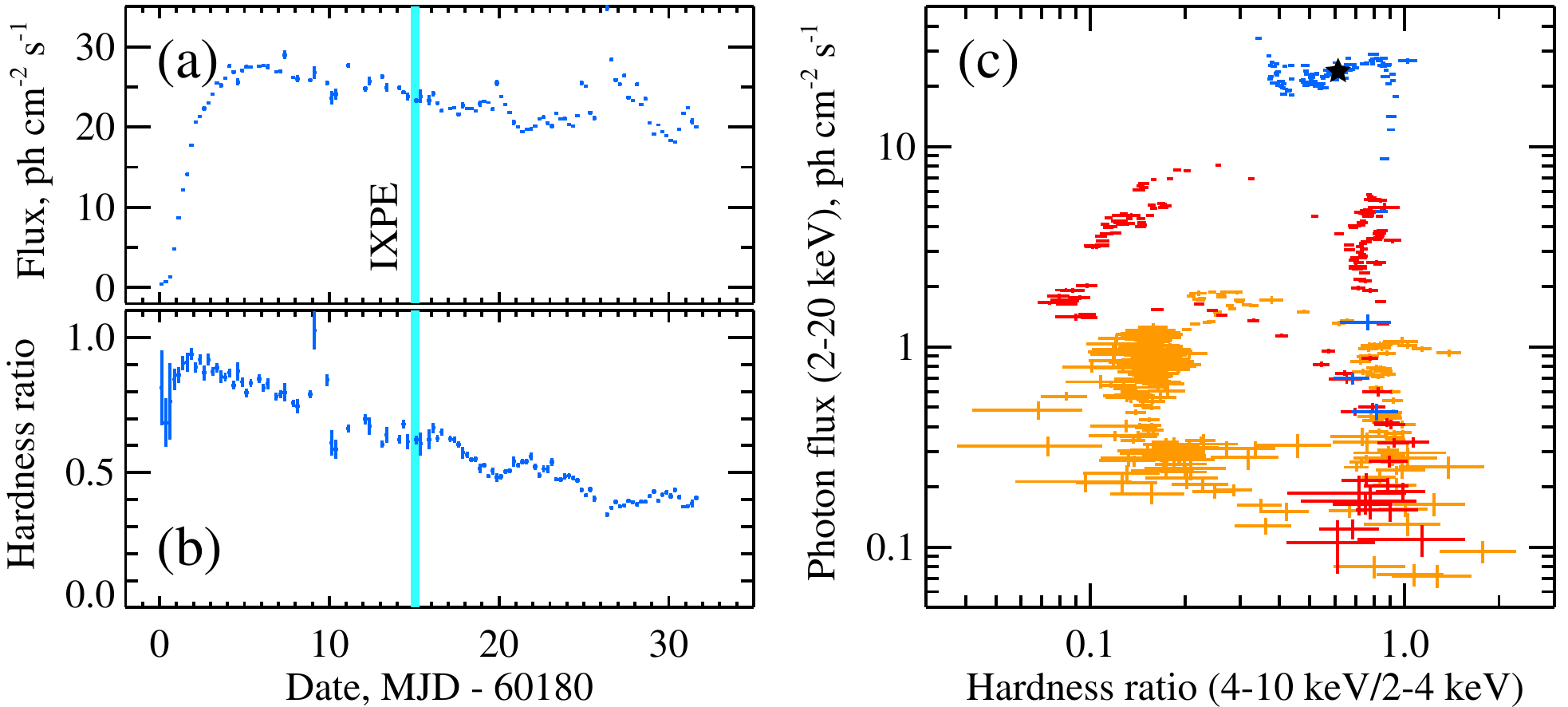}
\caption{Evolution of \source during the outburst. 
(a) MAXI light curve in the 2--20 keV range. The time of the IXPE observation is marked by the cyan vertical line.  
(b) Hardness ratio, i.e. the ratio of the photon flux in the 4--10 keV band to that in the 2--4 keV band. 
(c) Hardness-flux diagram. 
Blue crosses show the evolution of  \source during the current outburst. 
The position of the source on the diagram during IXPE observation is marked with a black star.  
For comparison, we also plot the time evolution of MAXI~J1820+070 during its 2018 outburst with red crosses and GX~339$-$4 during its 2010 outburst with the orange crosses. 
The diagram (c) shows that \source was observed in the hard state soon after the turning point.
} \label{fig:lc_hr}
\end{figure*}

The Imaging X-ray Polarimetry Explorer \citep[IXPE]{Weisskopf2022} provided the first significant detection of 2--8~keV X-ray polarization in the archetypical BH \mbox{Cyg~X-1} \citep{Krawczynski2022} -- the source that originally gave rise to the spectral classifications \citep{Tananbaum1972}.
The hot Comptonizing medium visible in the hard spectral state of the binary appeared to be elongated in the plane of the accretion disk.
The polarization angle (PA) of X-ray emission was found to be consistent with the position angle of the extended radio source \citep{Miller-Jones2021}, thus confirming, for the first time, that the radio jets are launched along the axis of the (inner) accretion disk.
The polarization degree (PD) of $4.0\%\pm0.2\%$ was higher than expected for the orbital inclination of the source $i\approx30\degr$ \citep{Miller-Jones2021}. 
This can be explained by the inner disk being more highly inclined than the outer disk \citep{Krawczynski2022} or by an outflow of the hot medium \citep{Poutanen2023}.

The BH X-ray binary \mbox{Cyg~X-3} became the second hard-state target of IXPE.
Similarly to \mbox{Cyg X-1}, the source swings between different spectral states, which might be attributed to the same configurations of accreting matter as in \mbox{Cyg~X-1}.
Contrary to early expectations, a high PD$=20.6\%\pm0.3\%$ was detected in the hard state \citep{Veledina2023}.
The X-ray PA was found to be shifted by $90\degr$ with respect to the position angle of the discrete radio ejections, as well as to the sub-mm polarization direction.
These unique polarization properties can be explained in terms of a pure reflection scenario.
The X-ray polarization observations revealed the presence of an obscuring and reflecting envelope at high elevation above the disk plane, suggesting this well-studied binary is a Galactic ultraluminous X-ray source.

The aforementioned hard-state sources belong to the class of high-mass X-ray binaries, where the compact object captures matter from the wind of their massive companion.
The innermost accretion geometry of low-mass X-ray binary BHs, where the disk forms via Roche lobe overflow, was not yet studied polarimetrically.
In this paper, we report on the first X-ray polarization measurement of the Galactic X-ray binary \source.

\section{Discovery and Outburst}
\label{sec:source}

\source underwent a bright outburst that was detected on 2023 Aug 24  \citep{GCN.34540,GCN.34544}.
MAXI \citep{Matsuoka2009} monitoring of the source showed it reaching $\sim$7~Crab in the 2--20~keV range (see Fig.~\ref{fig:lc_hr}a), triggering  interest and rapid initiation of follow-up multiwavelength campaigns by a number of ground-based and space observatories \citep{ATel16205,ATel16207,ATel16209,ATel16211,ATel16225,ATel16231}.
The beginning of the outburst was observed in the optical wavelengths about 4 days earlier than the first X-ray trigger \citep{ATel16209}.
Optical spectroscopy taken at the early outburst stages suggest the source is a low-mass X-ray binary \citep{ATel16208}, with signatures of an outflow \citep{ATel16216}.
On the next and the third day after the X-ray trigger, the X-ray spectral shape was identified to be consistent with the typical hard-state power law with photon index $\Gamma\sim1.3-1.7$, as measured by the LEIA imager \citep{ATel16210} and by the Mikhail Pavlinsky ART-XC telescope \citep{ATel16217}, respectively.
Recent NICER observations (taken 25 days after the beginning of the outburst) report substantial contribution of the blackbody emission (of temperature $kT_{\rm bb}=0.6$~keV) and softening of the spectrum \citep[$\Gamma=2.4$,][]{ATel16247}.
\mbox{X-ray} light-curves revealed prominent quasi-periodic oscillations with a slowly-increasing (on timescales of days) central frequency \citep{ATel16215,ATel16219,ATel16247}.

The onset of the radio source detected at various frequencies shortly after the X-ray trigger identified the presence of a jet \citep{ATel16211,ATel16228}, with a flat spectrum \citep[Trushkin et al., in prep.]{ATel16230} typical to the hard-state sources.
A subsequent sub-mm detection and polarization measurements indicate its direction is nearly along the North-South celestial axis \citep[PA=$-4\fdg1\pm3\fdg5$ on the date closest to IXPE pointing;][]{ATel16230}. 
The intrinsic optical polarization also seems to be roughly aligned with the North-South direction \citep{ATel16245}.
While the mass of the compact object has not been reliably measured yet, all indirect signatures indicate the binary hosts a BH, making it an intriguing target that can be used to probe the accretion geometry.

Fig.~\ref{fig:lc_hr}a shows the initial stages of the countrate evolution in \source as seen by MAXI monitor,\footnote{http://maxi.riken.jp/} and the typical fast rise profile can be clearly identified.
The starting point for the date corresponds to the initial X-ray trigger (which we will refer to as the beginning of the X-ray outburst hereafter).
A few days after the beginning of the outburst, the source started to soften (see Fig.~\ref{fig:lc_hr}b).
The source evolution follows the well-known q-track pattern \citep{Homan2005} in the hardness-flux diagram (blue symbols in Fig.~\ref{fig:lc_hr}c), albeit becoming much brighter than the prominent low-mass X-ray binary MAXI~J1820+070 (shown in red) and the prototypical source GX~339--4 (orange).
IXPE observed \source at high, close to current maxima, fluxes, about 15 days after the beginning of the outburst.
According to the position in hardness-flux diagram, the source started transition to the soft state and resided in the hard-intermediate state (which we refer to as the hard state hereafter, Fig.~\ref{fig:lc_hr}c).

\section{Observations and Data Reduction}
\label{sec:data}

IXPE is the first satellite mission dedicated to X-ray polarimetry in the 2--8 keV band \citep{Weisskopf2022}. 
It carries three X-ray telescopes, each made of a Mirror Module Assembly \citep{Ramsey2022} and a polarization-sensitive gas-pixel detector unit  \citep{Soffitta2021,Baldini2021}, that enable imaging X-ray polarimetry of extended sources and a huge increase of sensitivity for point-like sources.
IXPE provides an angular resolution of $\sim 30\arcsec$ (half-power diameter).
The overlap of the fields of view of the three detector units is circular with a diameter of 9\arcmin; spectral resolution is better than 20\% at 6~keV.

\begin{figure*}
\centering 
\includegraphics[width=0.75\textwidth]{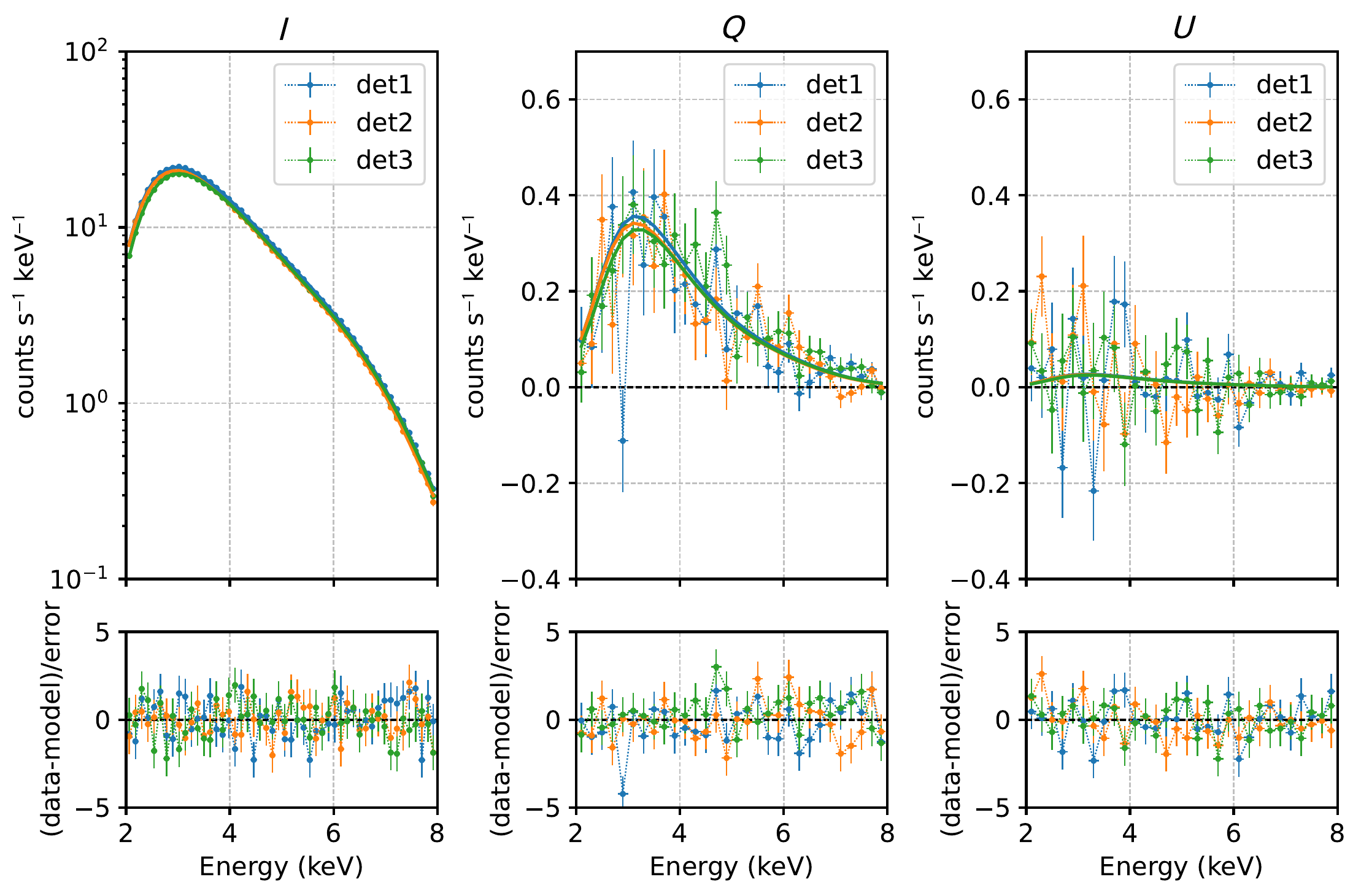}
\caption{IXPE X-ray spectra of  \source. 
Stokes $I$, $Q$, and $U$ spectra are shown in the left, middle, and right panels, respectively. 
Data from the three IXPE detectors are shown as crosses: det1 (blue), det2 (red), and det3 (green). 
The best-fit model \texttt{polconst*powerlaw} is shown with corresponding lines. 
The lower sub-panels show the fit residuals.
} \label{fig:ixpe_spectrum}
\end{figure*}

IXPE observed \source from 2023 Sept 7 19:59:07 UTC to Sept 8 06:35:12 UTC for a live\-time of $\sim$19~ks \citep{ATel16237,ATel16242}.
Level 2 data were downloaded from the IXPE archive at the HEASARC\footnote{https://heasarc.gsfc.nasa.gov/docs/ixpe/archive/} and analyzed with both \textsc{ixpeobssim} version 30.6.2 \citep{Baldini2022} and HEASOFT/\textsc{xspec} version 12.13.1d \citep{Arnaud1996}. 
IXPE observations of \source were carried out placing a ``gray'' filter in front of the detector  \citep{Ferrazzoli2020,Soffitta2021}. 
The filter was used to reduce (by a factor of $\sim$10) the incident flux from a very bright source, especially at low energies, to a level compatible with the dead-time of the detector units and with the need to transmit all data within the allocated telemetry.
Now different response matrices have to be used for data analysis, which are available both in the \textsc{ixpeobssim} package and in the HEASARC CALDB. 
Comparing the polarization obtained by the \textsc{pcube} algorithm in \textsc{ixpeobssim} and the one from \textsc{xspec}, we found that the latter provides more consistent results as it fully takes into account the energy response of the instrument. 
This is particularly important to correctly describe the low-energy response of the IXPE detectors when the gray filter is used, as its transmission drops steeply at low energies. 
In the following, we then present the results obtained with \textsc{xspec}. 
Source events were extracted using a circle with 80\arcsec\ radius centered on the source. 
No background subtraction is necessary for sources as bright as \source, since in this case the IXPE background is dominated by scattered source emission, even at large offset \citep[hence, the background itself is negligible and can be ignored,][]{DiMarco23}.

\begin{deluxetable*}{lccccccc}
\tablecaption{Energy dependence of polarization properties. 
\label{tab:pha1}}
\tablewidth{0pt}
\tablehead{
& 
\multicolumn{7}{c}{Energy range (keV) }\\  
\cline{2-8}
  \colhead{Parameter} & \colhead{2--3} & \colhead{3--4} & \colhead{4--5} &
\colhead{5--6} & \colhead{6--7} & \colhead{7--8}  & \colhead{{\bf 2--8}} 
}
\startdata
$Q/I$ (\%) & $3.4\pm0.5$ & $4.2\pm0.3$ & $4.2\pm0.4$ & $ 4.5\pm0.5$ & $5.1\pm0.8$  & $4.8\pm1.3$ & $4.1\pm0.2$\\
$U/I$ (\%) & $0.9\pm0.5$ & $0.3\pm0.3$ & $0.1\pm0.4$ & $0.0\pm0.5$  & $-0.7\pm0.8$ & $0.4\pm1.3$ & $0.3\pm0.2$\\
PD (\%)    & $3.5\pm0.5$ & $4.2\pm0.3$ & $4.2\pm0.4$ & $4.5\pm0.5$  & $5.1\pm0.8$  & $4.8\pm1.3$ & $4.1\pm0.2$\\
PA (deg)   & $7.4\pm3.7$ & $2.2\pm2.3$ & $0.7\pm2.7$ & $-0.2\pm3.4$ & $-3.7\pm4.2$ & $2.3\pm8.0$ & $2.2\pm1.3$ \\
\enddata
\tablecomments{Polarization characteristics are obtained from the $I$, $Q$, and $U$ Stokes parameters computed with \texttt{PHA1}, \texttt{PHA1Q}, and \texttt{PHA1U} algorithms in \texttt{xpbin} and fitted with \texttt{polconst*powerlaw} model in given energy intervals. 
The uncertainties are given at the 68.3\% (1$\sigma$) confidence level for one interesting parameter.  }
\end{deluxetable*}

\section{Polarization Results}
\label{sec:results}


We fitted Stokes $I$, $Q$, and $U$ spectra with \textsc{xspec}, using only simple models because of the limited spectral capability of IXPE. 
We first tested a \texttt{polconst*diskbb} model, representing an accretion disk consisting of multiple blackbody components with a constant polarization, using the \texttt{gain fit} command in \textsc{xspec} to fit the energy scale. 
We found an unacceptable fit, $\chi^2=3509$ for 1329 d.o.f., driven by large residuals in the $I$ spectrum. 
Models for absorption remain unconstrained due to the relatively high IXPE threshold (2~keV) and the use of the gray filter, which strongly reduces the flux from the source at low energies. 
Then, we used a power-law model assuming constant polarization \texttt{polconst*powerlaw}, using \texttt{gain fit} command.
The data with the best-fit model are shown in Fig.~\ref{fig:ixpe_spectrum}.
The model gives $\chi^2=1282$ for 1329 d.o.f. 
The best-fit power-law index is $\Gamma=1.80^{+0.02}_{-0.01}$, with the polarization being PD=$4.1\%\pm0.2\%$ at PA=$2\fdg2\pm1\fdg3$. 
The errors are at 68\% confidence level and calculated assuming one interesting parameter with the \texttt{steppar} command of \textsc{xspec}.
The polarization from the source is detected at a $\sim20\sigma$ confidence level.

\begin{figure}
\centering 
\includegraphics[width=0.47\textwidth]{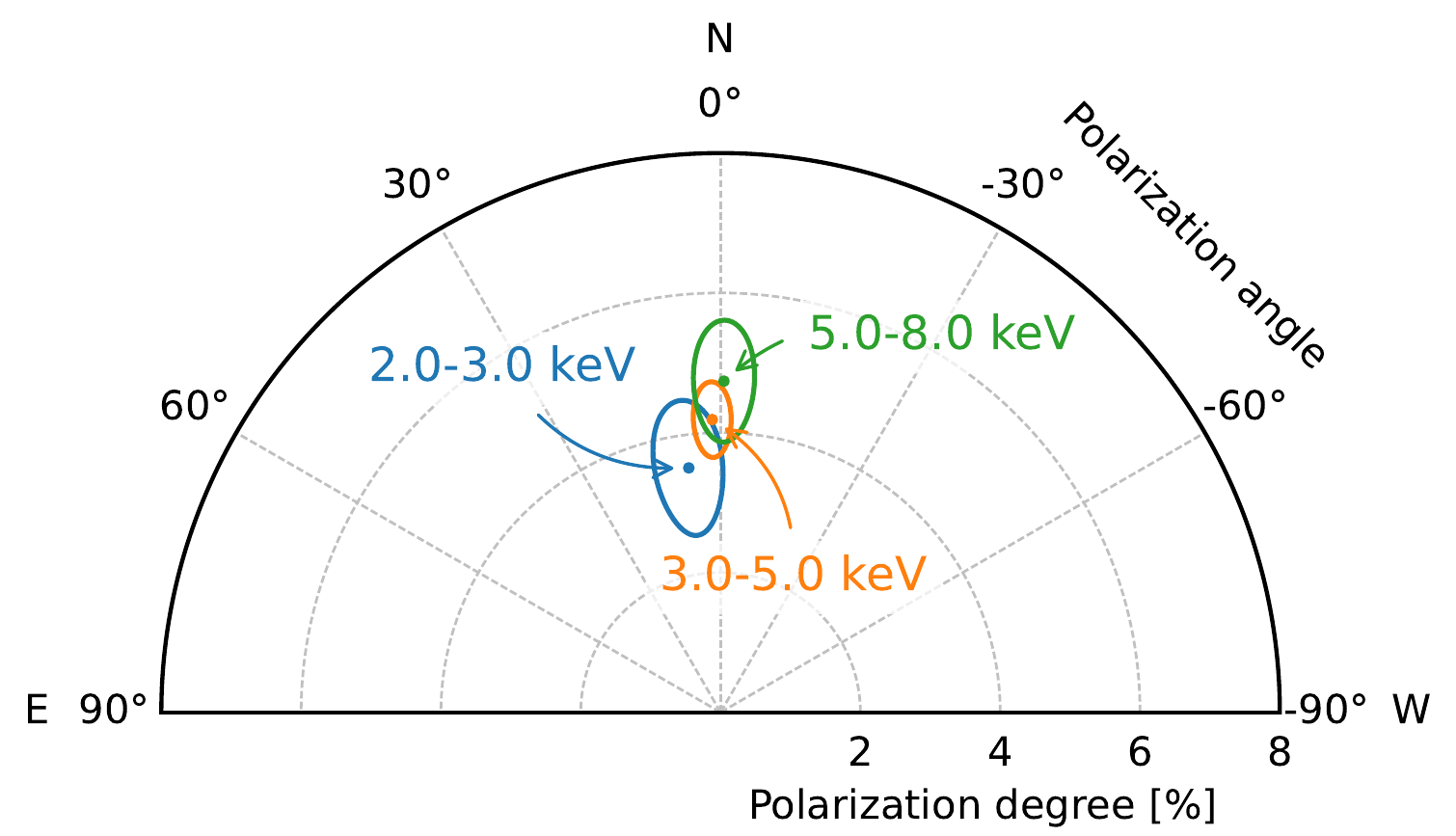}
\caption{Contour plots of the PD and PA in three  energy bands measured with \textsc{xspec} at 90\% confidence level.} \label{fig:pdpa}
\end{figure}

We then froze the spectral model to the one found for the 2--8 keV energy band and fitted in the different energy ranges leaving only the polarization model (either \texttt{polconst} or \texttt{stokesconst}) free to vary.
The results are given in Table~\ref{tab:pha1}.
The PD shows an increasing trend with energy, growing from about 3\% in the 2--3 keV band to $\sim$5\% above 5~keV. 
The PA does not show any significant variations with energy.  
Contours plots of the PD and PA in three energy bands are shown in Fig.~\ref{fig:pdpa}.

We then checked for a possible energy dependence of the polarization by substituting the \texttt{polconst} model with \texttt{pollin}, which assumes a linear dependence of PD and PA on photon energy. 
However, we assumed an energy-independent PA  ($\psi_1$ in \textsc{xspec}; fixing $\psi_{\rm slope}=0$), and allowed the PD to vary with photon energy $E$ (keV) as $\mbox{PD}(E)=A_1+A_{\rm slope}(E-1)$.  
We obtained a marginally better fit  $\chi^2/\mathrm{d.o.f.}=1277/1328$ with the F-test giving a probability of $3\%$ for a chance improvement. 
The best-fit parameters are $A_1=3.0\%\pm0.5\%$ and $A_{\rm slope}=0.34\pm0.13$\%\,keV$^{-1}$, and PA=$\psi_1=2\fdg0\pm1\fdg3$.

\section{Discussion}
\label{sec:discussion}

The exceptional brightness of the newly-discovered source \source is striking and can be compared to only a few historical BH transients: the low- or intermediate-mass X-ray binaries \mbox{A0620$-$00} \citep{Elvis1975}, \mbox{V404~Cyg} \citep{Zycki1999}, \mbox{V4641~Sgr} \citep{Revnivtsev2002}, \mbox{MAXI~J1820+070} \citep{Shidatsu2018}, and \mbox{4U~1543$-$47} \citep{Sanchez-Sierras2023}.
The high X-ray fluxes of \mbox{A0620$-$00} and \mbox{MAXI~J1820+070} are caused by the proximity of these sources to the observer, and the inferred accretion rates fall in the sub-Eddington category.
In these cases, we see the innermost parts of the disk/corona during the hard state.
On the other hand, the luminosities of \mbox{V404~Cyg},  \mbox{V4641~Sgr}, and \mbox{4U~1543$-$47} are believed to reach near- and super-Eddington values, and signatures of an optically thick outflow, that can cover the innermost regions, have been detected \citep{Revnivtsev2002,Motta2017,Prabhakar2023}.
Analogs to these classes of sources can be found in active galactic nuclei, whose geometry studies are now also enabled by IXPE data.
For the hard-spectrum sources where the innermost parts of the accretion disk are visible, the X-ray polarization is found to be in a range of about a few per cent (\mbox{Cyg~X-1}, \citealt{Krawczynski2022}; \mbox{NGC~4151}, \citealt{Gianolli2023}; \mbox{MCG-05-23-16}, \citealt{Marinucci2022,Tagliacozzo2023}; \mbox{IC~4329A}, \citealt{Ingram2023}) and aligned with the jet direction (wherever it is constrained).
For the sources where the central engine is obscured, the observed spectrum is dominated by a reflection continuum, and high (PD$>20$\%) polarization orthogonal to the jet direction (polarization aligned with the obscuring torus) has been found (Cyg X-3, \citealt{Veledina2023}; Circinus galaxy, \citealt{Ursini2023}).

The relatively low X-ray PD found in \source (as compared to some other accreting BHs) suggests the innermost parts of the accretion disk are visible, and the geometry is similar to that of \mbox{Cyg~X-1} and Seyfert~1 galaxies. 
This is supported by the hardness-flux diagram (Fig.~\ref{fig:lc_hr}c) which indicates that IXPE observed the source at the beginning of the hard to soft state transition. 
By comparing the flux of \source at the turning point (at which a source begins softening) with those of other BH X-ray binaries  \mbox{MAXI~J1820+070} situated at 3~kpc \citep{Atri20} and \mbox{GX~339$-$4} at an estimated distance of $\sim$10~kpc \citep{Zdziarski2019}, we can estimate that \source is closer than  \mbox{MAXI~J1820+070} by a factor of two (i.e. at a distance of about 1.5~kpc). 

Relativistic ejections have not yet been observed in the source, hence the direction of the jet is not known from radio images. 
The sub-mm polarization signal however gives a clue as to the jet axis \citep[oriented close to the North-South direction,][]{ATel16230}.
It has been found that the intrinsic polarization in the optical, sub-mm, and radio (after correction for Faraday rotation) is aligned with the jet direction in the X-ray binaries during the outburst (\mbox{Cyg~X-3}, McCollough et al. in prep.; Lange et al., in prep.; \citealt{Veledina2023};  \mbox{XTE~J1908+094},  \citealt{Curran2015};  \mbox{XTE~J1550$-$564}, \citealt{Migliori2017}; \mbox{MAXI~J1820+070}, \citealt{Veledina2019}; and \mbox{V404~Cyg}, \citealt{Kosenkov2017}).
If the jet direction in \source also aligns with the detected optical and sub-mm polarization, then the X-ray polarization is also aligned with the jet direction, supporting the analogy to sub-Eddington, rather than super-Eddington sources.

The similarity of the X-ray PA to the sub-mm PA and, by extension, to the position angle of the radio jet is related to the alignment of the jet direction with the BH spin and the jet collimation mechanisms \citep{McKinney2013,Davis2020}.
If the BH spin is misaligned with the binary orbital axis, the accretion disk/flow becomes warped and/or can experience Lense-Thirring precession \citep{Bardeen1975,Stella1998,Fragile2007}.
Precession of the accretion flow may then be visible in multiwavelength light-curves as prominent quasi-periodic oscillations \citep{Ingram2009,Veledina2013}, which are indeed detected in the X-ray timing data on \source \citep{ATel16215,ATel16219}.
We clearly detect two harmonics of these oscillations, at $1.340 \pm 0.004$ and $2.74 \pm 0.04$~Hz, in the IXPE flux data (Ewing et al., in prep.).
In the case of a substantial misalignment between the orbital and BH spin axes and rapid Lense-Thirring precession of the inner flow, the average (over the precession period) flow axis is aligned with the spin axis. 
As a consequence, the average X-ray PA is parallel to the BH spin axis.
The outer parts of the disk are, on the other hand, expected to be aligned with the orbital plane \citep[e.g., as found in MAXI~J1820+070,][]{Poutanen2022}.
The alignment of the jet direction with the BH spin in \source also suggests that the jets are launched from the radii that are either aligned with the BH spin axis or experience rapid precession around it.

The similarity of the detected X-ray PD in \source and \mbox{Cyg~X-1} may be interpreted as similarity of inclinations of the inner parts of the accretion disk in these systems, that translates to an intermediate inclination of the source $i\sim30\degr-60\degr$.
This is further supported by the absence of the X-ray dips.
The inclination of the binary orbit can be reliably measured only after the source reaches quiescence \citep{CasaresJonker2014}, and may help to distinguish between various configurations of the accretion disk and corona.
However, basic constraints on the geometry can be made using these first X-ray polarization measurements.
While the X-ray PA and its relation to the sub-mm PA suggest that the X-ray corona is extended along the disk (and not along the jet), the high PD$\sim$4\%, with some indication for increasing PDs with energy, argue against spherical and lamppost coronal geometries \citep{Krawczynski2022}.

\source continues to gradually soften, in line with the q-pattern in the hardness-flux diagram (Fig.~\ref{fig:lc_hr}c).
Subsequent observations by IXPE detecting the source in the soft and very high states may enable comparison to the other BH X-ray binaries observed in polarized X-ray light \citep{Ratheesh2023,RodriguezCavero2023,Podgorny2023,Svoboda2023,Dovciak2023}, providing independent constraints on the inclination and calibrating the models of accretion disk structure and radiative processes.

\section{Summary}
\label{sec:summary}

We obtained the first X-ray polarization measurement for the bright BH binary \source.
We find the time- and energy-averaged PD$= 4.1\%\pm0.2\%$ and PA$=2\fdg2\pm1\fdg3$ (68\% confidence) from \textsc{xspec} spectro-polarimetric fits.
There is no statistically significant dependence of the PA on energy. 
A model that assumes a linear dependence of PD on energy gives a slight improvement in $\chi^2$ significant at 97\% confidence level and results in a PD increasing with energy by 0.34\%~per~keV. 

The evolution of the source in the hardness-flux diagram and the observed spectral slope indicate that the source was in the hard state during IXPE observations, accreting at sub-Eddington rates, so its high X-ray brightness is caused by the source proximity to the Earth. 
We estimate the distance to the source to be about 1.5~kpc.

The alignment of X-ray, optical, and sub-mm polarization directions indicates that the hot medium producing the X-ray continuum emission is extended in the accretion disk plane (orthogonal to the jet).
Our findings support the expectation that the jets are launched orthogonal to the inner parts of the accretion flow.
By extension, this means that the direction of the inner parts of the jet in \source are aligned with the BH spin.

The PD value is comparable to that found in the high-mass BH X-ray binary  \mbox{Cyg~X-1}, and can, by similarity, constrain the inclination of the innermost parts to be moderate, $i\sim$30\degr--60\degr.
Also, the trend of PD with energy is inconsistent with the geometry of the spherical or lamppost corona, even if it is outflowing.
 
\section*{Acknowledgments}
The Imaging X-ray Polarimetry Explorer (IXPE) is a joint US and Italian mission.
The US contribution is supported by the National Aeronautics and Space Administration (NASA) and led and managed by its Marshall Space Flight Center (MSFC), with industry partner Ball Aerospace (contract NNM15AA18C). 
The Italian contribution is supported by the Italian Space Agency (Agenzia Spaziale Italiana, ASI) through contract ASI-OHBI-2017-12-I.0, agreements ASI-INAF-2017-12-H0 and ASI-INFN-2017.13-H0, and its Space Science Data Center (SSDC), and by the Istituto Nazionale di Astrofisica (INAF) and the Istituto Nazionale di Fisica Nucleare (INFN) in Italy.
This research has made use of the MAXI data provided by RIKEN, JAXA and the MAXI team.

A.V. thanks the Academy of Finland grant 355672 for support.
M.D., J.S., J.Pod., and V.Kar. thank GACR project 21-06825X for the support and institutional support from RVO:67985815. 
A.I. acknowledges support from the Royal Society. 
H.K. acknowledges support by NASA grants 80NSSC22K1291, 80NSSC23K1041, and 80NSSC20K0329.
The French contribution is supported by the French Space Agency (Centre National d'Etude Spatiale, CNES) and by the High Energy National Programme (PNHE) of the Centre National de la Recherche Scientifique (CNRS).
I.L. was supported by the NASA Postdoctoral Program at the Marshall Space Flight Center, administered by Oak Ridge Associated Universities under contract with NASA.


%

\vspace{5mm}
\facilities{IXPE, MAXI} 



\software {\textsc{ixpeobssim} \citep{Baldini2022}, \textsc{xspec} \citep{Arnaud1996}
}

\bibliographystyle{aasjournal}


\end{document}